# This is a shell archive.  Remove anything before this line, then
# unpack it by saving it in a file and typing "sh file".  (Files
# unpacked will be owned by you and have default permissions.)
#
# This archive contains:
# preprint.tex figures.ps

echo x - preprint.tex
cat > "preprint.tex" << '//E*O*F preprint.tex//'
\documentstyle[11pt]{article}
\newcommand{\ignore}[1]{}

\begin{document}
\baselineskip 16pt
\thispagestyle{empty}

\bigskip
\bigskip

{\center \Large \bf Virtual Bond Percolation for \\
             Ising Cluster Dynamics\\}

\bigskip
\bigskip

\bigskip
{\center

R. Brower\\
\smallskip
{\it
Department of Physics, Boston University, \\
590 Commonwealth Avenue, Boston, MA 02215}\\

\bigskip

P. Tamayo\\
\smallskip
{\it
Thinking Machines Corp.\\
245 First. St. Cambridge, MA 02142}\\}

\bigskip
\bigskip
\bigskip
\bigskip

{\center \today\\}

\bigskip
\bigskip

{\center \bf Abstract\\}
\bigskip
The Fortuin-Kasteleyn mapping between the Ising model and the site-bond
correlated percolation model is shown to be only one of an infinite class of
exact mappings. These new cluster representations are a result of
``renormalized'' percolation rules correlated to entire blocks of spins.  For
example these rules allow for percolation on ``virtual'' bonds between spins
not present in the underlying Hamiltonian. As a consequence we can define new
random cluster theories each with its own Monte Carlo cluster dynamics that
exactly reproduce the Ising model. By tuning parameters on the critical
percolation surface, it is demonstrated numerically that cluster algorithms
can be formulated for the 2-d and 3-d Ising model that have smaller
autocorrelations than the original Swendsen-Wang algorithm.

\newpage

\bigskip

{\bf 1. Introduction}

Our understanding of the Ising model has been greatly enhanced by the study of
Coniglio-Klein \cite {CK} critical percolation clusters and the classic work
of Fortuin-Kasteleyn\cite{FK,PERCO} on the exact mapping of the Ising model
into the random cluster model for site-bond correlated percolation. With the
work of Swendsen and Wang \cite{SW} this cluster representation has also lead
to
accelerated Monte Carlo dynamics with very small autocorrelation times
\cite{BROWER1,EDWARDS,KLEIN}.

In all of these endeavors the underlying clusters representation appeared to
be unique --- the probability of bond percolation, the only free parameter,
being fixed by detailed balance or the requirement of scaling in the critical
region.  However, a central mystery of this beautiful formalism remains
concerning the incompletely decoupling of clusters as manifest in divergent
autocorrelations at the phase transition.  It appears to us that a ``perfect''
representation of normal modes via clusters (analogous to fourier modes for a
pure Gaussian theory) should in principle exhibit no critical slowing down (z
= 0).

Here we show that the underlying Fortuin-Kasteleyn map is not unique.

By introducing longer range correlated percolation, we show that there are
additional free parameters (in the null vector space transverse to the
critical percolation surface) which are not fixed by the requirement of an
exact mapping between the Ising and the random cluster partition functions.
The percolation process is extended to longer length scales without leaving
the critical surface on which the pair connectedness function coincides
identically with the Ising correlation function.  The new freedom to define
``renormalized'' critical clusters may help to understand the residual cluster
interactions and lead to even more efficient Monte Carlo methods. By removing
the rigid constraints of earlier methods, we also hope to formulate new
cluster methods for more complex systems such as spin glasses\cite{GLASS}.
Even generalizations to continuous molecular systems now appear less daunting.

In the present article our new cluster methods will be explained exclusively
in the context of the Ising model, although it should be obvious that they
apply equally to all the recent extensions of the Swendsen-Wang algorithm such
as those made by Brower and Tamayo\cite{BROWER1} for the
Landau-Ginzburg theory, by Wolff\cite{WOLFF} for spin models and others
\cite{EDWARDS,KLEIN,GLASS}. While shorter autocorrelations evidently can
result,
in all probability the dynamical
universality class and the critical exponent $z$ will not be changed unless
we go to recursive methods that introduce virtual bonds on all length
scales.  Such extensions of our methods will be considered in future
publications.

The organization of this article is as follows. Section 2 presents the block
percolation scheme through two simple examples, with illustrations of possible
extensions. Section 3 defines the resulting  cluster dynamics and computes
autocorrelations and cluster distributions. The renormalized clusters are
shown numerically to lead to reduced autocorrelations relative to the
Swendsen-Wang method.  Appendix A presents the general construction of our
block correlated decomposition and the proof of equilibrium for the resultant
cluster dynamics. Although the general argument is important both for its
rigor and for the extraordinary range of alternative cluster methods it
implies, the specific examples are probably easier to understand.

\bigskip
{\bf 2. Blocking Schemes for the 2-d and 3-d Ising Model}

Let us begin by studying in detail the simplest extension of the
Fortuin-Kasteleyn (F-K) mechanism from blocks of 2 spins to blocks of 3 spins.
This modest generalization already encounters several qualitatively new
features.

The conventional F-K mapping has two steps.
The first step decomposes the Ising Hamiltonian into a sum over bonds (or 2
spin blocks)
so that the Ising probability distribution is seen as a product of factors.
\begin{equation}
P^{Ising}(s_i) = { 1 \over Z} \prod_{<i,j>} e^{ - \beta(1 - s_i s_j)}
\end{equation}
The second step decomposes each factor into two terms,
\begin{equation}
e^{ -\beta(1 - s_1 s_2)} = \sum_{n = 0, 1} p_n \; \Delta_n(s_1, s_2) = e^{ - 2
\beta} +  (1 - e^{-2\beta}) \delta_{s_1, s_2}
\label{FKDEC}
\end{equation}
with the probabilities,
\begin{equation}
p_0 =  e^{ - 2 \beta}  \; , \;\;\; p_1 =  1 - e^{-2 \beta}  \; ,
\end{equation}
uniquely determined by the requirement that probabilities for aligned {\sl vs}
unaligned spins be correct.  The deletion (n=0) and percolation (n=1) terms
have been designated by $\Delta_0(s_1, s_2) = 1$ and $\Delta_1(s_1, s_2) =
\delta_{s_1, s_2} $ respectively to accommodate
a more general expansion for larger blocks.

{\bf 2.1 Three Spin Blocks}

The general expansion for a block of spins B takes the form,
\begin{equation}
e^{- \beta V_B(s_i)} = \sum_{n = 0,..., n_{max}-1} p_n \; \Delta_n(s_i).
\end{equation}
Each of the $n$ terms is a unique product of delta functions
representing one of the many possible ways to break it up into clusters.  The
only requirement is that the entire Hamiltonian be given by the sum over the
set of blocks and that the expansion of the block into delta functions be
complete.  (See Appendix A for the proof and an even larger class of options.)

For example the 3-spin block for a nearest neighbor Ising lattice (See Figure
1 and 2), $B \equiv (s_1,s_2, s_3)$, can be split into 5 terms,
\begin{equation}
e^{-\beta(2 - s_1 s_2 - s_2 s_3)}\\ = p_0 + p_1 \; \delta_{s_2,s_3}
+ p_2 \; \delta_{s_1,s_3}   + p_3 \; \delta_{s_1,s_2} + p_4 \; \delta_{s_1,s_2}
\;
\delta_{s_2,s_3}
\end{equation}
However this most general 3-spin expansion\footnote{Note that the
identity,  $\delta_{s_1,s_2} \;
\delta_{s_2,s_3} = \delta_{s_1,s_2} \; \delta_{s_1,s_3} \;
\delta_{s_2,s_3}$, implies that the last term is actually a symmetric
cluster of all three spins.} is {\bf no longer unique}. Since there are 5
independent $p's$ and only 4 independent equations, for $(s_1, s_2, s_3)$ =
$(+,+,+)$, $(-,+,+)$, $(+,-,+)$, $(+,+,-)$, the linear equations must have a
non-trivial null space. The general solution, $p_i = p^{FK}_i + p^{null}_i$,
is the sum of an original inhomogeneous F-K solution $p^{FK}_i$ and one null
vector, $p_i^{null} = (-1, 1, 1, 1, -2)$.
\begin{eqnarray}
p_0 &=& e^{- 4\beta} - \zeta \nonumber \\
p_1 &=& e^{-2\beta} (1 - e^{- 2\beta})  + \zeta \nonumber \\
p_2 &=&  \zeta \nonumber \\
p_3 &=& (1 - e^{- 2\beta}) e^{-2\beta} + \zeta \nonumber \\
p_4 &=& (1 - e^{- 2\beta})^2 - 2 \zeta \;,
\end{eqnarray}
This null vector expresses the simple fact that at least two Ising spins must
be in the same state.
\begin{equation}
(1 -  \delta_{s_1,s_2} ) (1 - \delta_{s_2,s_3}) ( 1 -  \delta_{s_3,s_1}) = 0
\label{nonanalytic}
\end{equation}
No new solutions are present for the q state Potts model for $q > 2$ without
going to larger blocks.  The parameter $\zeta$ has been chosen to represent a
virtual bond connecting $s_1$ to $s_3$ so that at $\zeta = 0$ this
solution reverts to the standard Fortuin-Kasteleyn decomposition. As the
parameter $\zeta$ is increased the ``real'' bonds are reduced and the longer
range ``virtual'' bond increased moving along the critical surface of the
mapping.  The remarkable simplicity of percolation relative to most critical
phenomena is seen in the ability to find the critical surface analytically as
a function of the critical Ising temperature.

For Monte Carlo cluster algorithms, it will turn out that the only requirement
is to make all the probabilities non negative, which in the neighborhood of
the critical temperature restricts $\zeta$ to $ 0 \le \zeta \le e^{-
4\beta_c}$.  For the 2-d Ising at the critical temperature, the limiting case,
$\zeta = e^{- 4\beta_c} = (1 - \sqrt 2)^2$ has the remarkable property that
both the no percolation term $p_0$, and the fully percolated term $p_4$, are
exactly zero.  Consequently any 3 spins must have precisely one percolated
bond with 100\% probability.

To illustrate a little the range of options we have considered a  variety of
ways of partitioning the 2-d lattice into 3 spin (or two bond) blocks (See
Figure 1). It is instructive to consider the limiting case $\zeta = e^{-
4\beta}$ for lattice 1a with the three spin blocks centered on all the ``black
sites''. If we start with a checkerboard and use our limiting case again,
$\zeta = e^{- 4\beta_c}$, the lattice will immediately percolate into two
sublattices with infinite clusters on the ``red'' sites and single clusters on
the ``black'' sites. The study of such extreme non-equilibrium configurations
shows how different our clusters are relative to Coniglio-Klein site-bond
correlated percolation clusters.

It  is easy to extend this example to the most general
three spin blocks,
\begin{equation}
e^{ -\beta_{12}(1 - s_1 s_2) -\beta_{23}(1 - s_2 s_3)-\beta_{31}(1 - s_3 s_1)
}\\ = p_0 + p_1 \; \delta_{s_2,s_3} + p_2 \; \delta_{s_1,s_3} + p_3 \;
\delta_{s_1,s_2} + p_4 \; \delta_{s_1,s_2} \;
\delta_{s_2,s_3} \label{triang}
\end{equation}
appropriate for continuous spin models on cubic or triangular lattices.  Again
the general solution can be expressed as $p_i = p^{FK}_i + p^{null}_i$, where
the inhomogeneous solution is given by the F-K prescription applied
to each bond separately and the null vector is unchanged. If we consider
a red-black decomposition of the triangular lattice ($\beta_{ij} = \beta$), the
limiting case , $\zeta = e^{- 6\beta_c} = \sqrt3^3$ at the critical point
again has the remarkable property that one and only one bond is percolated
with 100\% probability.

In the remainder of the paper we will consider in greater detail the resultant
random cluster model and its Monte Carlo dynamics for the 3-spin block and
plaquette blocking schemes. However as the reader will readily note, we have
opened a Pandora box of alternative schemes. As the blocks are increased in
size, there are more and more free parameters and neither the blocks nor the
choice of these parameters have to be chosen to respect the symmetries of the
underlying lattice.  Consequently the {\bf exact} constraint of mapping the
Ising model into a random cluster scheme is far less restrictive than we might
expected.

{\bf  2.2  Plaquette Blocking for the Ising Model.}

As a second additional example we briefly describe a scheme
corresponding to 4 spin blocks or plaquettes.  The energy can be expanded
either, as a sum over all plaquettes dividing the interaction $\beta$ for
each bond by the number of ways it is shared (2 in 2-d or 4 in 3-d), or by
breaking the lattice up into a sum over over red (or black) plaquettes on
the lattice. As in the 3-spin block case the Boltzmann factor for each
plaquette is decomposed in a weighted sum over products of delta functions
$\Delta_n$ that constrain different subsets of spins inside the plaquette.
\bigskip
\bigskip
\begin{eqnarray}
e^{- \beta V_{plaq}} & = & p_0 + p_1 \; ( \delta_{s_1, s_2} + \delta_{s_3, s_4}
+
\delta_{s_1, s_3} + \delta_{s_3, s_4}) + \nonumber\\ & & p_2 \; ( \delta_{s_3,
s_1} \delta_{s_1, s_2} + \delta_{s_1, s_2} \delta_{s_2, s_4} + \delta_{s_2,
s_4}
\delta_{s_4,s_3} + \delta_{s_4, s_3} \delta_{s_3, s_1} ) + \nonumber\\
& & p_3 \; ( \delta_{s_1, s_3} \delta_{s_2, s_4} + \delta_{s_1, s_2}
\delta_{s_3,
s_4} ) + p_4 \; \delta_{s_1, s_2} \delta_{s_2,s_4} \delta_{s_4, s_3}
\delta_{s_3, s_1} +
\nonumber\\ & & p_5 \; \delta_{s_1, s_4} \delta_{s_2, s_3} + p_6 \;
(\delta_{s_1,
s_4} + \delta_{s_2, s_3}),
\end{eqnarray}
where $s_1, s_2, s_3$ and $s_4$ are the four spins that define the
plaquette (see figure 3). Notice that the probabilities $p_5$
and $p_6$ represent virtual percolation bonds across the diagonals that
freeze spins which do not interact in the Ising Hamiltonian.

This expansion is not the most general however, since we have imposed
rotational symmetry in the spins ($s_1, s_2, s_3$, $s_4$). For example the use
of
three spin blocks as shown in figure 1d offers a legitimate example of a four
spin expansion as well, which violates rotational symmetries on the plaquette.
The most general four spin expansion has 15 parameters and only 7 constraints.
The number of solutions grows faster than exponential as the size of the block
increases --- for a 5 spin block there are already 52 parameters but only 16
constraints. Our 5 spin star blocking (fig 1a) could be expanded with 36
adjustable parameters !

Imposing rotational symmetry for the Ising model, there are four
independent linear equations for $(s_1,s_2, s_3, s_4) = (+,+,+,+), (-,+,+,+),
(-,-,+,+), (-,+,-,+)$ and their solution is expressed again as the sum of the
F-K inhomogeneous solution plus the null space --- this time with three
free parameters $(\zeta_1, \zeta_2,
\zeta_3)$ for three null vectors.
\begin{eqnarray}
p_0 & = & e^{- 8 \beta} - 4 \zeta_2 - 2 \zeta_3\nonumber\\
p_1 & = &  e^{- 6 \beta}( 1  - e^{- 2\beta}) +  \zeta_1 +  3
 \zeta_2 + \zeta_3\nonumber\\
p_2 & = &  e^{- 4 \beta} (1 - e^{- 2 \beta})^2 - 2  \zeta_1 -  2  \zeta_2 -
\zeta_3 \nonumber\\
p_3 & = &  e^{- 4 \beta} (1 - e^{- 2 \beta})^2   - 2  \zeta_1 -  2
\zeta_2\nonumber\\
p_4 & = &4  e^{- 2 \beta} (1 - e^{- 2 \beta})^3 + (1 - e^{- 2 \beta})^4 +  8
\zeta_1\nonumber\\
p_5 & = & 4 \zeta_2\nonumber\\
p_6 & = & \zeta_3   \label{plaq}
\end{eqnarray}
The parameters $\zeta's$ are constrained by the condition that all
probabilities must be positive numbers in the range (0, 1). These three
parameters give us additional flexibility to tune the percolation process in
different ways. For example, one can increase or decrease the probability of
diagonal bonds with respect to horizontal or vertical bonds and, as we will
see in section 5, in this way change cluster properties and decorrelation
times. The plaquette example can also be applied to the 3 state Potts model
with the result that the rotational symmetric solutions has only one null
vector, $p^{null}_i = (1,-1,2,1,-6,1,-1)$, expressing the fact that in a 3
state
Potts model the four spins of a plaquette can all reside in different states.
In general it appears likely that for a q-state Potts model one must resort to
blocks bigger than q spins to find a non-trivial null space.

Finally we would like to emphasize that rather elegant blocking schemes
exist whereby bonds are shared between adjacent blocks with the energy divided
between the two blocks. Even for the F-K map this is allowed due to the
trivial identity on each shared bond that $((1-p)\delta_{s_1,s_2} + p) \times
( (1-p')\delta_{s_1,s_2} + p') = (1-p p')\delta_{s_1,s_2} + p p'$. For example
in our blocking schemes, we can restore translational symmetry for the 2-d
triangular and square lattices simply by the replacement $\beta \rightarrow {1
\over 2} \beta$ throughout eq. \ref{triang} and eq. \ref{plaq}
respectively.  Each bond is split into two equal pieces.
For the Plaquette percolation method this has the appealing feature that all
the symmetries of the original theory are preserved with next-to-nearest
neighbor virtual bond percolation allowed over the entire lattice. We feel
that the more symmetric forms are more likely to express the proper scaling of
critical clusters and thus lead to more efficient algorithms.

\bigskip
{\bf 3. Cluster Dynamics. }

In order to define a dynamics based on the percolation mappings described in
the previous section one has to introduce for every blocking scheme, a set of
auxiliary percolation variables $n_B = 0, 1, ...$ on each block. Then the
Monte Carlo cluster dynamics is chosen to be a Markov process for the joint
distribution $P_B^{joint}(s_i, n_B)$, in which one alternatively chooses the
percolation variables $n_B$ at fixed spins $s_i$ and the spins for each
cluster at fixed $n_B$ (see Appendix A).

We outline our algorithm in the following steps.
\begin{enumerate}
\item Choose a local blocking scheme in such way that all spin-spin
interactions
are accounted for in the ensemble of blocks. Expand each block into a fixed
percolation representation to define the joint distribution, $P_B^{joint}(s_i,
n_B)$.
\item For each block obtain its spin configuration ($s_i$)
and choose a value for the ``percolation'' variable $n \equiv n_B$
with normalized probability:
	\begin{equation}
	p(n|s_i) = \frac{p_n \Delta_n(s_i)}{\sum_n p_n \Delta_n(s_i)}
	\end{equation}
\item  Label the connected components for the percolation graph using
a cluster finding algorithm such as breadth-first  search\cite{SEARCH}, the
Hoshen-Kopelman algorithm\cite{HK}, or a parallel algorithm \cite{MG}.
\item  Flip all of the spins in each cluster coherently with $50$ \%
probability and return to step 2.
\end{enumerate}

Basically this is very similar to the standard Swendsen-Wang algorithm except
that the percolation step involves probabilities correlated to entire blocks
of spins. The variations on Swendsen-Wang suggested by Wolff\cite{WOLFF} can
be adapted to these clusters as well.  There is a slight increase in
complexity in the percolation step compared with Swendsen-Wang but the
algorithm is still very straightforward.  The values for the $\zeta's$
parameters can be taken to have different values for different block in the
system. In our simulations we choose all of them to be fixed at a uniform
value and then generate a look-up table of probabilities to be employed in the
percolation process. The 3-spin and 4-spin block dynamics at $\zeta_i = 0$ are
identical to the original Swendsen-Wang algorithm.

\bigskip
{\bf 3.1 Numerical Results}

Due to their simplicity we have implemented the 3-spin and plaquette blocking
schemes as described in the previous section. 	In this section we analyze the
autocorrelation times for equilibrium simulations and the dynamics of cluster
distributions as the system relaxes to equilibrium.

{\em Autocorrelation times analysis}

All our simulations were carried out at the critical point of the infinite
system. We studied the autocorrelation function for the squared
magnetization as an average over Monte Carlo time ($2.5 \times 10^5$ steps).
Figure 4 shows two autocorrelation functions for a system of size $64
\times 64$ using the 3-spin block dynamics with the star blocking shown in
figure 1a.  The two curves correspond to the values of $\zeta$ at the two
extremes of the interval $(0,\;\; e^{-4
\beta_c})$ of allowed values.  The two functions show exponential relaxation
but the extreme case farthest from the Swendsen-Wang point is faster by a
noticeable amount.  Moreover, the data in figure 5 appears to exhibit a
monotonic decrease in autocorrelation times as the probability of virtual
bonds increases. A similar picture appears for the 3-spin block dynamics in
three dimensions where for $\zeta = e^{-4 \beta_c}$, $\tau_{m^2}$ is almost
twice as fast as Swendsen-Wang.

For the red-black plaquette dynamics we explored different regions of the
solution space. We have not done an exhaustive search but it is clear that
the autocorrelation times decrease as $\zeta_1$ decreases and the other two
parameters, $\zeta_2$ and $\zeta_3$, increase. The fastest dynamics, the one
with smallest
autocorrelation times, we were able to find is the one corresponding to
$\zeta_1 = -0.056349, \;
\zeta_2 = 0.00245, \; \zeta_3 = 0.0098$ which is about 1.6 times faster
than the Swendsen-Wang dynamics. Similar increases in speed are obtained for
the
symmetric form which expands both red and black plaquettes simultaneously
using eq. 10 with $\beta \rightarrow { 1 \over 2}\beta$.

A general trend is seen in all these cases, as one might expect, increasing
the probability of long range virtual bonds reduces the
autocorrelation times.

Figure 6 shows the scaling of autocorrelation times with system size for the
3-spin block, plaquettes and Swendsen-Wang dynamics in two dimensions. If
indeed the scaling is the same for all these dynamics, then they will have the
same dynamic exponent $z$ as the original Swendsen-Wang and consequently would
belong to the same dynamic universality class\cite{SW,HH}.  A more
comprehensive and systematic study of these dynamics is required to settle
this question but all our numerical results appear to be consistent with
universality.

{\em Cluster relaxation analysis}

In addition to the standard autocorrelation times analysis we studied the
equilibrium cluster distributions and the dynamics of cluster relaxation (see
Stauffer, Kertesz and Miranda \cite{CLUSTERS}).  Figure 7 shows the
equilibrium cluster distributions for a $32 \times 32$ system using the
plaquette blocking and Swendsen-Wang.  Notice that the two cluster
distributions are not identical but have very similar scaling behavior. The
fastest plaquette dynamics having larger probabilities for virtual bonds
favors larger clusters. It is clear that the cluster distributions follow a
power law with identical or similar exponents.  These exponents are related to
the fractal dimension, $d_f$, of the clusters by the relation, $n^{eq}_s \sim
s^{-(1 + d/d_f)}$ with the implication that our extended percolation clusters
have the same or similar fractal dimension as the Coniglio-Klein clusters.

In order to study cluster relaxation, a large number of simulations were
performed in which the system relaxes to equilibrium at the critical point
{}from an initial state with all spins up.  Cluster histograms were
accumulated at every time step and a logarithmic binning used to reduce
fluctuations.  Bin $n_s$ contains contributions from clusters sizes in the
interval $(2^s, 2^{s+1} -1)$.  Clusters numbers are assumed to relax
exponentially to equilibrium,
\begin{equation}
n_s(t) \sim (1 - e^{-t/\tau_s}) \; n^{eq}_s.
\end{equation}
The scaling of the cluster decorrelation time $\tau_s$ with cluster size
defines a critical exponent $r$,
\begin{equation}
\tau_s \sim s^r \sim s^{z/d_f}
\end{equation}
where $z$ is the standard critical slowing down exponent and $d_f$ is the
fractal
dimension of the clusters $(d_f = d - \beta/\nu)$. For the 2-d Ising model,
$d_f = d - \beta/\nu  = 1.875$.


We have performed this analysis for two values of $\zeta$ in the 3-spin block
dynamics. We used a system of size $256^2$ and sampled over 400 experiments.
Figures 8 shows the relaxation of $n_s(t)$ as a function of time for the
Swendsen-Wang and the $\zeta=e^{-4 \beta_c}$ dynamics. It is clear that the
$\zeta=e^{-4 \beta_c}$ dynamics produces a much more rapid relaxation process
for the cluster numbers, perhaps by a factor of 5. This is surprising in view
of the much smaller change for the equilibrium autocorrelation time (a factor
of about 1.5), and gives an indication that the non-equilibrium dynamics may
depend very sensitively on the strength of virtual bonds.  In fact, the
relaxation is so fast that it is hard to extract the scaling behavior by
computing relaxation times for different cluster numbers. A more thorough
study of the effects of these dynamics on cluster properties would help us to
understand how to tune the $\zeta$ parameters to have more control over the
cluster relaxation process.

All the numerical experiments confirm the validity of these dynamics as
equilibrium Monte Carlo processes and their potential to reduce
autocorrelation times.

\bigskip
{\bf 4. Conclusions}

The main purpose of this article was to demonstrate the existence of a new
class of exact percolation mappings not allowed in the original methods of
Fortuin and Kasteleyn.  To clarify the discussion we have studied in detail 3
spin and 4 spin blocking schemes applied to the nearest neighbor Ising model
on a cubic lattice. However as we demonstrate in Appendix A, the freedom
to invent new cluster algorithms within this framework is extraordinarily
unencumbered. Not only is the Fortuin-Kasteleyn mapping not unique, one can
find {\bf exact} mappings which have peculiar properties such as not
respecting the exact symmetries of the underlying model. We also note as
illustrated by eq. \ref{nonanalytic} that these new mappings can not in general
be extended analytically to the entire q state Potts model. The lower the value
of q the larger the class of solutions.  Now the problem is no longer
the difficulty of finding percolation mappings but the problem of choosing the
most interesting or efficient straw in the haystack.

It is difficult not to be reminded of the analogous freedom in blocking
schemes for the real space Renormalization Group (RG). In spite of this
analogy several differences remain.  For example the real space
Renormalization Group has the following properties not shared by our approach:
(i) the Hamiltonian has fewer degrees of freedom so that only the longer
length scales are described correctly, (ii) the blocking procedure gives
renormalized parameters that are automatically tuned to respect the scale
change and (iii) the blocking is applied recursively. With respect to the
first feature concerning the use of a semi-group, perhaps our approach offers
an improvement in that we have an exact isomorphism. However the second two
features we would like to emulate more closely, in the belief that they
might be the key to better cluster methods.  For example, it
appears likely to us that the optimal choices for our $\zeta_i$ parameters (ie
those with shortest autocorrelations) can be ``derived'' from scaling
arguments. We are also looking at recursive blocking schemes --- in particular
a class of recursive algorithms based on iterative block decomposition in the
spirit of the Migdal-Kadanoff renormalization group, which may be able to
reduce $z$ below the value given by the Swendsen-Wang multi-cluster or Wolff
single-cluster algorithms. Progress on these issues will be published in the
future.

It is also interesting to consider the possibilities that virtual bond
percolation parameters may be able to tune percolation processes in cases in
which the standard Swendsen-Wang percolation does not produce acceleration
such as in spin-glasses.  Work is also proceeding on this problem. In summary
the use of virtual bond dynamics in its most general form may be able to open
new perspectives for cluster methods both to understand critical domains or
droplets and to design faster algorithms for Monte Carlo simulations in
statistical mechanics and quantum field theory.


{\bf Acknowledgments}

We would like to thank R. Giles, W Klein, H. Gould, S. Huang, J. Mesirov,
G. Batrouni and B. Boghosian, for interesting discussions. We also thank
D.  Stauffer for suggesting to us the cluster analysis of section 3.1.

 \newpage
 \centerline {\bf Figure Captions}
 \begin{itemize}
 \item {Figure 1. Examples of the variety of blocking schemes for
 3-spins. The blocking used in our numerical experiments is shown in a). This
 blocking can be seen as a special case of a 5-spin ``star'' blocking.  It is
 also interesting to notice that c) is an asymmetrical blocking and that d)
 produces plaquette-like structures.}
 \item {Figure 2. The expansion of the interaction between 3-spins in terms of
 percolation variables.}
 \item {Figure 3. Different percolation schemes for the 4-spin plaquette block
mapping.}
 \item {Figure 4. Magnetization squared autocorrelation functions for the
 3-spin block dynamics in two (a) and three (b) dimensions with different
 parameters $\zeta$. The two values shown correspond to the extremes of the
 interval of allowed values.}
 \item {Figure 5. Variation of the magnetization squared autocorrelation time
as a
 function of $\zeta$ for the 3-spin block dynamics in two dimensions. The value
 of $\tau_{m^2}$ decreases as $\zeta$, the probability of virtual bonds,
increases. }
 \item {Figure 6. Scaling of $\tau_{m^2}$ as a function of system size $L$
 for the 3-spin block, plaquettes and Swendsen-Wang dynamics in two
 dimensions.}
 \item {Figure 7. Equilibrium cluster distributions for Swendsen-Wang and
 plaquette dynamics (with $\zeta_1 = -0.056349, \zeta_2 = 0.00245$ and $\zeta_3
=
 0.0098$) in two dimensions. The slopes are nearly identical but large
 clusters are favored by the plaquette dynamics.}
 \item {Figure 8. Relaxation of cluster numbers for Swendsen-Wang (a) and the
 3-spin block (for $\zeta = e^{-4 \beta_c}$) (b) dynamics in two dimensions.}
 \end{itemize}

\newpage

\setcounter{equation}{0}
\renewcommand{\theequation}{A.\arabic{equation}}

{\bf Appendix A: General Formulation of Block Percolation}

Our generalization of the Fortuin-Kasteleyn percolation map has a strong
similarity to blocking methods used for the real space Renormalization Group
transformations. To bring out this similarity and to prove the correctness of
the resulting dynamics, we follow the suggestion of Kandel et al\cite{GLASS} by
introducing effective block potentials, $U^{eff}_B(s_i,n_B,\beta)$, instead of
the delta function constraints $\Delta_{n_B}(s_i)$ considered above.  Also we
note that the delta functions can be recovered as a limiting case. For
example, in this approach the original F-K map is replaced by the following
two terms,
\begin{equation}
e^{ -\beta(1- s_1 s_2)}= p_0 e^{ -J_0 (1 - s_1 s_2)} + p_1 e^{ -J_1 (1 - s_1
s_2)}
\end{equation}
where $U^{eff}_B(s_i,n,\beta) = J_n (1 - s_1 s_2) + log(p_n)$. If we
adjust the two weights correctly,
\begin{eqnarray}
p_0 &=& (e^{-2 \beta} - e^{- 2 J_1})/ (e^{-2 J_0} - e^{- 2J_1})
\;,\nonumber\\
p_1 &=& (e^{-2 J_0} - e^{- 2 \beta})/(e^{-2 J_0} - e^{- 2 J_1}) \;,
\end{eqnarray}
this is in itself a (new) legitimate mapping of the Ising bond at temperature
$\beta^{-1}$ into two randomly chosen terms with effective temperatures
$J_i^{-1}$.
We can recover the standard F-K map by taking the limit of the effective
{\it inverse} temperatures to zero ($J_0 \rightarrow 0$) and to  infinity ($J_1
\rightarrow \infty$) for the deletion  and  freezing (ie percolation)
terms respectively (see eq.\ref{FKDEC}).  It is also interesting to view
percolation
methods in this way as the limiting extreme of a random mapping of a fixed
temperature Ising system into a heterogeneous mixture of high and low
temperature interactions.

Our general blocking scheme has two steps:
\begin{enumerate}
\item The Hamiltonian is expressed as a sum over blocks $B$ of spins,
\begin{equation}
H(s_i) = \sum_B V_B(s_i).
\end{equation}
\item The corresponding factor in the  partition function is decomposed
into a sum of terms,
\begin{equation}
e^{- \beta V_B(s_i)} = \sum_{n_B= 0,1,...n_{max} - 1} e^{-U^{eff}_B(s_i, n_B,
\beta)}
\label{factor}
\end{equation}
\end{enumerate}
The products over all the resulting terms (\ref{factor}) forms a new joint
probability distribution, $P^{joint}(s_i, n_B)$ so that our generalized map is
\begin{equation}
P^{Ising}(s_i) = \sum_{\{n_B\}} P^{joint}(s_i, n_B),  \label{Ising:marginal}
\end{equation}
The Ising model is represented as a marginal distribution by summing over
the generalized percolation variables, $n_B$. Alternatively the sum over the
Ising variables yields the generalized ``random cluster'' (RC) model,
\begin{equation}
P^{RC}(n_B) = \sum_{\{s_i\}} P^{joint}(s_i, n_B).  \label{RC:marginal}
\end{equation}
Lastly by taking the limiting case for effective potentials to freeze or
delete clusters of spins in each block, we can among others obtain the entire
class of virtual bond (or extended) percolation models.  The fundamental
identity for this class of percolation models, that guarantees criticality of
the random cluster model at the Ising fixed point, is
\begin{equation}
\langle s_i s_j \rangle \equiv \sum_{\{s_k\}} s_i s_j P^{Ising}(s_k) =
\sum_{\{n_B\}}
\gamma_{i,j} P^{RC}(n_B)
\end{equation}
where $\langle s_i s_j  \rangle$ is the Ising correlation function and the
connectedness
function, $\gamma_{i,j}$, is 1 if $i$ and $j$ are in the same random cluster
and 0 otherwise. From this identity it follows that the Ising spin-spin
correlation length and the cluster connectedness length are equal and
therefore that  the critical scaling of both the Ising model and the
corresponding
random cluster model are identical.

{\em Equilibrium for cluster Monte Carlo}

To prove that all static quantities are given correctly by the Monte Carlo
cluster algorithm in Section 3, one must show that the Markov process
preserves the equilibrium distribution,
\begin{equation}
P^{Ising}(s'_i) =  \sum_{\{s_i\}} W(s'_i \leftarrow s_i) \; P^{Ising}(s_i).
\label{balance}
\end{equation}
This equilibrium {\it balance} condition is weaker than the full {\it
detailed } balance satisfied by many Monte Carlo algorithms, but it is
sufficient. To prove it recall that in our cluster Monte Carlo algorithm the
percolation variables $n_B$ in each block were first chosen according to the
conditional probability,
\begin{equation}
p(n_B|s_i) = \frac{ e^{-U^{eff}_B(s_i, n_B, \beta)} }{ \sum_{n_B} e^{-
U^{eff}_B(s_i, n_B, \beta)} } = e^{- ( U^{eff}_B(s_i, n_B, \beta) - \beta
V_B(s_i))}
\end{equation}
and ``clusters'' of spins were subsequently chosen at fixed values of $n_B$.
Clearly this implies that the full transition matrix is a product of two
conditional distributions,
\begin{equation}
W(s'_i \leftarrow s_i)  = \sum_{\{n'_B\}} P^{joint}(s_i|n'_B)
P^{joint}(n'_B|s_i).
\label{wequ}
\end{equation}
To complete the last step in demonstration one explicitly performs the sums in
eq. \ref{wequ} and eq. \ref{balance} by applying Bayes theorem,
\begin{equation}
P(a,b) = P(a|b) P(b) \;\;\; P(b) \equiv \sum_b P(a,b),
\end{equation}
for each of the marginal distributions in eq. \ref{Ising:marginal} and eq.
\ref{RC:marginal}. While there is nothing fundamentally new in this proof
relative to that needed for the Swendsen-Wang algorithm it is necessary to see
that it still goes through. Moreover this more general derivation illuminates
a striking feature of cluster methods that the stochastic replacement of the
full potential by a particular effective potential bears a very close
resemblance to the Metropolis algorithm. Put into the context of blocks of
spins, the similarity to the multigrid concept of ``stochastic coarsening'' is
also transparent.


\begin{thebibliography}{999}

\bibitem{CK} A. Coniglio and W. Klein,
{\sl J. Phys. A}, {\bf 13}, 2775 (1980).

\bibitem{FK} C. M. Fortuin and P. W. Kasteleyn,
{\sl Physica}, {\bf 57}, 536 (1972); P. W. Kasteleyn and C. M.
Fortuin, {\sl J. Phys. Soc. Japan Suppl.} {\bf 26s}, 11 (1969).

\bibitem{PERCO} D. Stauffer,
{\sl Introduction to Percolation Theory}, Taylor and Francis, London (1985).

\bibitem{SW} R. Swendsen and J. S. Wang,
{\sl Phys. Rev. Lett.}, {\bf 58}, 86 (1987).

\bibitem{BROWER1} R. Brower and P. Tamayo,
{\sl Phys. Rev. Lett.}, {\bf 62}, 1087 (1989).

\bibitem{EDWARDS} R. Edwards  and A. Sokal,  {\sl Phys. Rev. D}
{\bf 38}, 2009 (1988); R. G. Edwards and A. D. Sokal,
{\sl Phys. Rev. D}, {\bf 38}, 2009 (1988); A. Sokal, ``Monte Carlo
Methods in Statistical Mechanics: Foundations and New Algorithms'',
Lecture Notes.

\bibitem{KLEIN} R. Brower and S. Huang, {\sl Phys. Rev. D,} {\bf 41},
 708 (1990); P. Tamayo, R. C. Brower and W. Klein, {\sl J. of Stat.
Phys.}, {\bf 58}, 1083 (1990); E. Marinari and R. Marra, ROM2F-89-41 (Rome
preprint 1989); H. G. Evertz, R. Ben-Av, M. Marcu and S. Solomon,
FSU-SCRI-90-196 (1990). F. Niedermayer {\sl Phys. Rev. Lett.} {\bf 61},
2026 (1988); T.  Ray, P. Tamayo and W. Klein, {\sl Phys.  Rev. A} {\bf 39},
5949 (1989).

\bibitem{GLASS} D. Kandel, R. Ben-Av and E. Domany, {\sl Phys. Rev. Lett.}
{\bf 65}, 941 (1990); D. Kandel and E. Domany, {\sl Phys. Rev. B} {\bf 43},
8539 (1991).

\bibitem{WOLFF}  U. Wolff, {\sl Phys. Rev. Lett.} {\bf 60}, 1461 (1988).

\bibitem{SEARCH} T. H. Cormen, C. E. Leiserson and R. L. Rivest, {\sl
Introduction to Algorithms}, MIT Press, (1990); D. Knuth, {\sl The Art of
Computer Programming, vol 3}, Addison-Wesley (1973).

\bibitem{HK} J. Hoshen and R. Kopelman, {\sl Phys. Rev. B}, {\bf 14}, 3438
(1976).

\bibitem{MG} R. C. Brower, Pablo Tamayo and Bryant York, {\sl Jour. of Stat.
Phys.} {\bf 63} 73 (1991); M. Flanigan and P. Tamayo (in preparation); J.
Apostolakis, P. Coddington and E. Marinari, {\sl Europhys. Lett.} {\bf 17}
189 (1992).  C. F. Baillie and P. D. Coddington, {\sl Concurrency: practice
and Experience}, {\bf 3(2)}, 129 (1991); A. N. Burkitt and D. W. Heermann,
{\sl Comp. Phys. Comm.} {\bf 54}, 210 (1989).

\bibitem{HH}P. C. Hohenberg and B. Halperin, {\sl Rev. of Mod. Phys.}, {\bf
49}, 435 (1977).

\bibitem{CLUSTERS}
D. Stauffer, {\sl Physica A} {\bf 171}, 471 (1991); D. Stauffer and J.
Kertesz, {\sl Physica A} {\bf 177}, 381 (1991); E. N. Miranda, {\sl Physica
A} {\bf 175}, 229 (1991), {\bf 175}, 235 (1991), {\bf 179}, 340 (1991).


\end{thebibliography}
\end{document}